\documentstyle[amssymb,prl,aps,epsfig]{revtex}

\begin{document}

\title{Evidence for a smooth superconductor to normal state transition for nonzero
applied magnetic field in Sr$_{0.9}$La$_{0.1}$CuO$_{2}$}

\author{T. Schneider$^{\text{1}}$, R. Khasanov$^{\text{1,2,3}}$, D. Di Castro$^{\text{1,4}}$, H. Keller$^{\text{1}}$, Mun-Seog Kim $^{\text{5,8}}$,
C. U. Jung $^{\text{6,8}}$, Sung-Ik Lee $^{\text{7,8}}$}
\address{$^{\text{(1)}}$ Physik-Institut der Universit\"{a}t Z\"{u}rich, Winterthurerstrasse 190, CH-8057, Switzerland\\
$^{\text{(2)}}$ Laboratory for Neutron Scattering, ETH Z\"{u}rich and PSI Villigen, CH-5232 Villigen PSI, Switzerland\\
$^{\text{(3)}}$DPMC, Universit\'e de Gen\`eve, 24 Quai
Ernest-Ansermet, 1211 Gen\`eve 4,
Switzerland\\
$^{\text{(4)}}$ INFM-Coherentia and Dipartimento di Fisica,
Universita' di Roma "La Sapienza", P.le A. Moro 2, I-00185 Roma,
Italy
\\$^{\text{(5)}}$Division of Electromagnetic Metrology,
Korea Research Institute of Standards and Science\\ P. O. BOX 102,
Yuseong, Daejeon 305-600, Korea\\$^{\text{(6)}}$Department of
Physics, Hankuk University of Foreign Studies, Yongin, Kyungki
449-791, Korea\\$^{\text{(7)}}$Quantum Materials Research
Laboratory, Korea Basic Science Institute, Daejeon 305-333,
Korea\\$^{\text{(8)}}$National Creative Research Initiative Center
for Superconductivity, POSTECH, Pohang 790-784, Korea}

\maketitle

\begin{abstract}
The effect of the magnetic field on the critical behavior of
Sr$_{0.9}$La$_{0.1}$CuO$_{2}$ is investigated near the zero field
superconductor to insulator transition at $T_{c}$. We present and
analyze field cooled magnetization data, revealing for $0\leq \mu
_{0}H\lesssim 5$ T remarkable consistency with a magnetic field
induced finite size effect. It is traced back to the fact that the
correlation length $\xi $ cannot grow beyond the limiting length
scale $L_{H}$ set by the magnetic field, where at temperature
$T_{p}\left( H\right) $, $\xi \left( T\right) =L_{H}$. Thus, in
sufficiently homogeneous samples and nonzero $H$ the transition
from the superconducting to the normal state turns out to be
smooth and the appropriately scaled magnetization data fall near
$T_{p}\left( H\right) $ on a universal curve. Consistent with the
generic behavior of optimally doped cuprates we also show that the
pressure effect on $T_{c}$ is negligibly small, while the negative
value of the relative volume change under pressure mirrors that of
the anisotropy.
\end{abstract}
\bigskip

In this study we present and analyze magnetization data of the
infinite-layer compound Sr$_{0.9}$La$_{0.1}$CuO$_{2}$. Since near
the zero field transition thermal fluctuations are expected to
dominate\cite{jhts,book,parks} and in sufficiently high fields
these fluctuations become effectively one dimensional \cite{lee},
whereupon the effect of fluctuations increases with increasing
magnetic field, it appears unavoidable to account for thermal
fluctuations. Indeed, invoking the scaling theory of critical
phenomena it is shown that the data are inconsistent with the
traditional mean-field interpretation. On the contrary, we observe
agreement with a magnetic field induced finite size effect.
Indeed, when the magnetic field increases, the density of vortex
lines becomes greater, but this cannot continue indefinitely, the
limit is roughly set on the proximity of vortex lines by the
overlapping of their cores. Because of the resulting limiting
length scale $L_{H}$ for a field applied parallel to the c-axis
the correlation length $\xi _{ab}$ cannot grow
beyond\cite{parks,haussmann,lortz,bled}
\begin{equation}
L_{H}=\sqrt{\Phi _{0}/aH},  \label{eq1}
\end{equation}
with $a\simeq 3.12$\cite{bled}. It is comparable to the average
distance between vortex lines and implies that $\xi $ cannot grow
beyond $L_{H}$ and with that there is a magnetic field induced
finite size effect. This implies that thermodynamic quantities
like the magnetization, magnetic penetration depth, specific heat
\textit{etc.} are smooth functions of temperature near
$T_{p}\left( H\right) $, where the correlation length cannot grow
beyond $\xi \left( T_{p}\right) =L_{H}$. This scenario holds true
when the magnetization data $M\left( T,H\right) $ collapses near
$T_{p}\left( H\right) $ on a single curve when plotted as
$M/(TH^{1/2})$ \textit{vs}. $\left( T/T_{c}-1\right) /\left(
1-T_{p}\left( H\right) /T_{c}\right) $. We observe that the
magnetization data falls for $0\leq \mu _{0}H\lesssim 5$ T within
experimental error on a single curve by adjusting $T_{p}\left(
H\right) $. From the resulting field dependence of $T_{p}$ we
deduce for the critical amplitude of the correlation length the
estimate $\xi _{ab0}\simeq 35$\AA . Noting that $L_{H}$ decreases
with increasing field strength it sets unavoidably the limiting
length scale for sufficiently high fields, the effect of the
inhomogeneity induced counterpart can be eliminated. In the
present case, both the magnetic field induced shift of the peak
location and peak height in $dM/dT$ clearly reveal that for $\mu
_{0}H\gtrsim 0.1$T the magnetic field induced finite size effect
allows to probe the homogeneous parts of the grains with dimension
$\gtrsim 814$\AA . The observed consistency with the magnetic
field induced finite size effect then implies that whenever the
thermal fluctuation dominated regime is accessible, sufficiently
homogeneous type II superconductors do not undergo a continuous
phase transition in a nonzero magnetic field, \textit{e.g.} to a
state with zero resistance. Furthermore, we show that the pressure
effect on the magnetization does not provide estimates for the
shift of the transition temperature only, but involves the change
of the volume and anisotropy as well, in analogy to
MgB$_{2}$\cite{tsdc}. Consistent with the generic behavior of
optimally doped cuprates we find that in
Sr$_{0.9}$La$_{0.1}$CuO$_{2}$ the pressure effect on $T_{c}$ is
negligibly small, while the negative value of the relative volume
change mirrors that of the anisotropy.

\begin{figure}[tbp]
\centering
\includegraphics[totalheight=6cm]{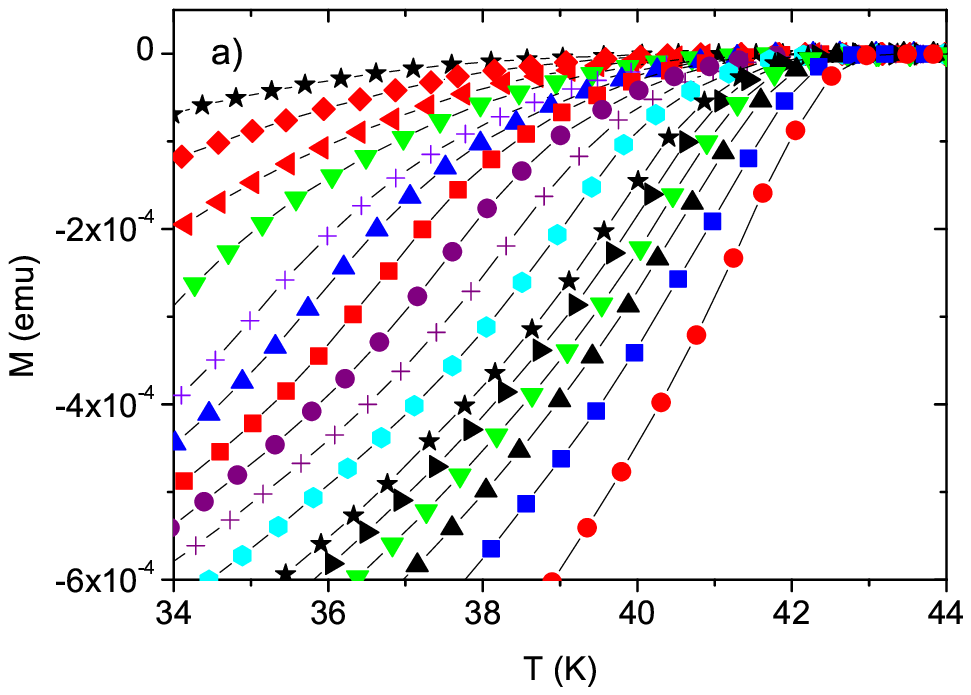}
\includegraphics[totalheight=6cm]{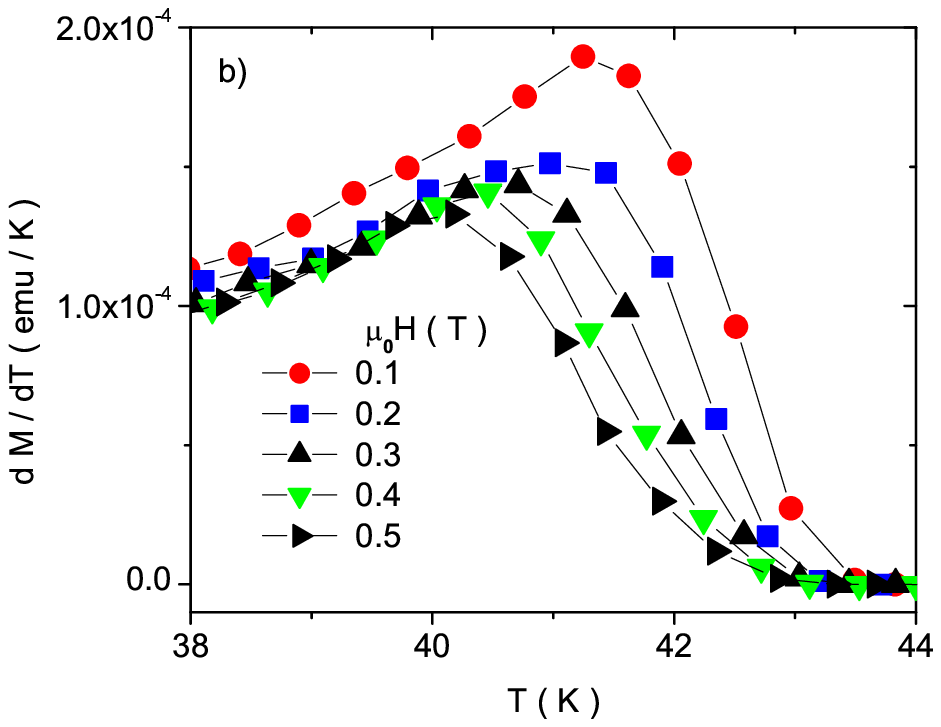}
\caption{a) Field cooled magnetization $M$ of a
Sr$_{0.9}$La$_{0.1}$CuO$_{2}$ powder sample \textit{vs}. $T$ for
various applied magnetic fields. From the left to the right: $\mu
_{0}H=5$, $4$, $3$, $2.5$, $1.75$, $1.5$, $1.25$, $1$, $0.8$,
$0.6$, $0.5$, $0.4$, $0.3$, $0.2$, and $0.1$ T; b) $dM/dT$
\textit{vs.} $T$ for some data shown in Fig.\ref{fig1}a.}
\label{fig1}
\end{figure}

\begin{figure}[tbp]
\centering
\includegraphics[totalheight=6cm]{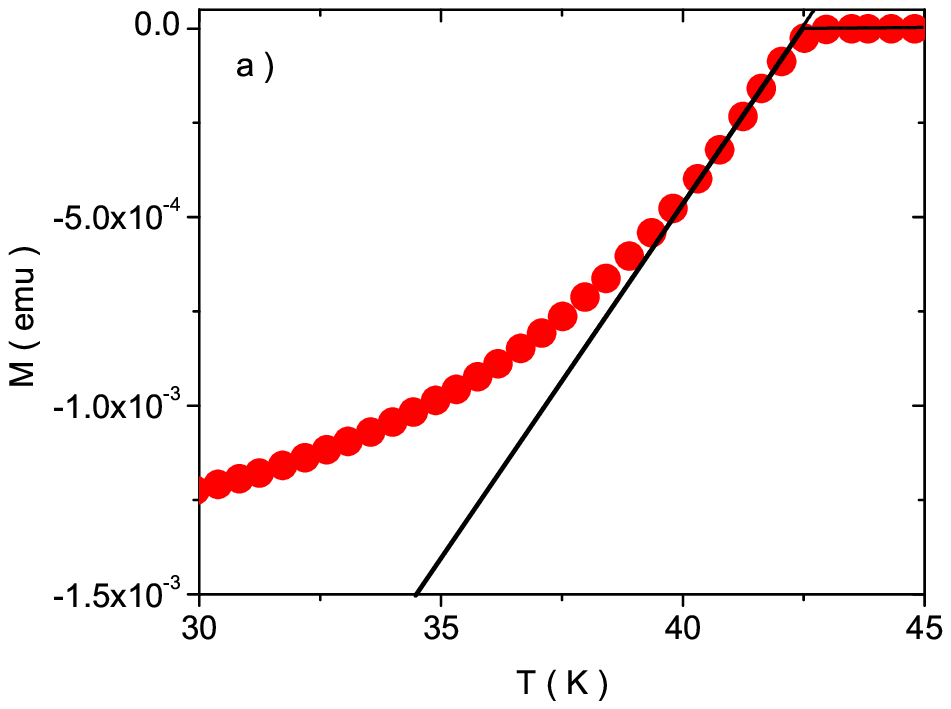}
\includegraphics[totalheight=6cm]{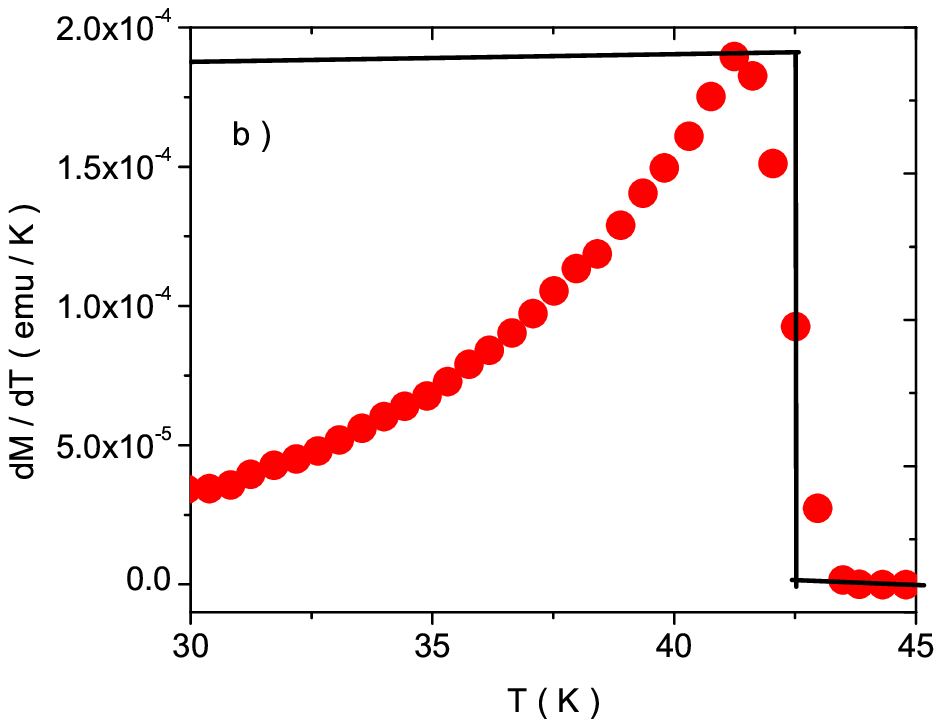}
\caption{a) $M$ vs. $T$ for $\mu _{0}H=0.1$T $\left( \bullet
\right) $ for the respective data shown in Fig.\ref{fig1}a; the
solid line presumes the linear slope given by Eq.(\ref{eq2}) with
$T_{c}=42.4$K; b) $dM/dT$ vs. $T$ for $\mu _{0}H=0.1$T $\left(
\bullet \right) $ for the data shown in Fig.\ref {fig2}a; the
solid lines indicate the discontinuous behavior resulting from
Eq.(\ref{eq3}).} \label{fig2}
\end{figure}

In Fig.\ref{fig1}a we displayed our ac-field-cooled magnetization
data in terms of $M$ \textit{vs.} $T$. Some respective $dM/dT$
\textit{vs.} $T$ are shown in Fig.\ref{fig1}b. For a detailed
description of the sample preparation we refer to Kim \textit{et
al}.\cite{kim}. The ac-field-cooled magnetization measurements
were performed with a Quantum Design (PPMS) magnetometer at
temperatures ranging from $5$ to $50$K. The ac-frequency was set
to $1000$Hz and the amplitude to $0.5$mT. To identify the
temperature regime where critical fluctuations play an essential
role it is instructive to compare the generic features of the data
with the predictions of Abrikosov's mean-field
treatment\cite{abrikosov} whereupon near the upper critical field
$H_{c2}$ the magnetization is given by
\begin{equation}
4\pi M\left( T\right) =-\frac{1}{\left( 2\kappa ^{2}-1\right)
\beta _{A}}\left( H_{c2}\left( T\right) -H\right) ,  \label{eq2}
\end{equation}
and in turn
\begin{equation}
4\pi \frac{dM\left( T\right) }{dT}=-\frac{1}{\left( 2\kappa
^{2}-1\right) \beta _{A}}\frac{dH_{c2}\left( T\right) }{dT},\text{
}\frac{dH_{c2}\left( T\right) }{dT}=-\frac{\Phi _{0}}{2\pi \xi
_{0}^{2}T_{c2}\left( H\right) }. \label{eq3}
\end{equation}
$\kappa =\lambda /\xi $ denotes the Ginzburg-Landau parameter,
$\beta _{A}=1.16$ for a hexagonal vortex lattice, $\xi =\xi
_{0}\left( 1-T/T_{c2}\left( H\right) \right) ^{-1/2}$ the
correlation length and $T_{c2}\left( H\right) $ the mean-field
transition temperature at the upper critical field critical field
$H_{c2}\left( T\right) $. Thus, this approximation predicts a
continuous phase transition in an applied field along the line
$H_{c2}\left( T\right) =\Phi _{0}/\left( 2\pi \xi _{0}^{2}\right)
\left( 1-T/T_{c0}\right) $, where $T_{c0}$ is the zero field
mean-field transition temperature. In Fig.\ref{fig2} we compare
the essential predictions with the magnetization data for $\mu
_{0}H=0.1$T. Although the magnetization appears to be consistent
with a linear slope below some temperature ''$T_{c2}\left(
H\right) $'' and becomes very small above, the data for $dM/dT$
deviates from the resulting discontinuity. According to
Fig.\ref{fig1}b this holds true for higher fields as well.
Furthermore, the absence of a field dependence of $dM/dT$ below
''$T_{c2}\left( H\right) $'' is not confirmed either. In
principle, the rounding of the transition could be due to a
inhomogeneity or/and magnetic field induced finite size effect, or
the variation of superconducting properties from grain to grain.
However, since the rounding of the transition increases with the
applied magnetic field (Fig.\ref{fig1}b) and the magnetic field
induced limiting length scale $L_{H}$ (Eq.(\ref{eq1})) decreases
one suspects that the rounding stems from the magnetic field
induced finite size effect. In this case there is no sharp
superconductor to normal state transition in an applied magnetic
field, as predicted by the mean-field approximation. To verify
this educated guess we take thermal fluctuations, the magnetic
field and the inhomogeneity induced finite size effect into
account.

\bigskip

When the rounding of the transition stems from a magnetic field or
inhomogeneity induced finite size effect, the correlation length
$\xi $ cannot grow beyond the limiting length $L_{H,I}$,
where\cite{jhts,book,parks,bled,tsjh}
\begin{equation}
\xi \left( T_{p}\right) =\xi _{0}\left| t_{p}\right| ^{-\nu
}=L_{H,I},\text{ }t=1-T_{p}/T_{c},\nu \simeq 2/3.  \label{eq4}
\end{equation}
$L_{I}$ denotes the limiting length of the homogeneous domains of
the sample. Note that $\nu \simeq 2/3$ holds in the charged and
uncharged (3D-XY) universality class \cite{book,tsrkhk}.In
superconductors, exposed to a magnetic field $H$ , there is the
aforementioned additional limiting length scale $L_{H}=\sqrt{\Phi
_{0}/\left( aH\right) }$(Eq.(\ref{eq1})), related to the average
distance between vortex lines. Indeed, as the magnetic field
increases, the density of vortex lines becomes greater, but this
cannot continue indefinitely, the limit is roughly set on the
proximity of vortex lines by the overlapping of their cores.
Because of these limiting length scales the phase transition is
rounded and occurs smoothly. Consequently, the thermodynamic
quantities like the magnetization, magnetic penetration depth,
specific heat \textit{etc.} are smooth functions of temperature
near $T_{p}$. To uncover the scaling properties of the
magnetization in this regime and to estimate the magnetic field
dependence of $T_{p}$, we invoke the scaling properties of the
free energy per unit volume in the fluctuation dominated regime.
For a field applied with an angle $\delta $ from the c-axis it
reads\cite{jhts,book,parks,bled,tsjh}
\begin{equation}
f=\frac{Qk_{B}T}{\xi _{ab}^{2}\xi _{c}}G\left( z\right)
,~z=\frac{H\xi _{ab}^{2}\epsilon }{\Phi _{0}},\epsilon =\left(
\cos ^{2}\left( \delta \right) +\frac{1}{\gamma ^{2}}\sin
^{2}\left( \delta \right) \right) ^{1/2}, \label{eq5}
\end{equation}
where $Q$ is a universal constant, $G\left( z\right) $ a universal
function of its argument, $\xi _{ab,c}$ the correlation lengths
parallel to the $ab$- and $c$-axis, respectively, and $\gamma =\xi
_{ab}/\xi _{c}$ the anisotropy. For the magnetization per unit
volume we obtain then the scaling relation
\begin{equation}
m=-\frac{\partial f}{\partial H}=-\frac{Qk_{B}TH^{1/2}}{\Phi
_{0}^{3/2}}\gamma \epsilon ^{3/2}\frac{1}{z^{1/2}}\frac{dG}{dz}.
\label{eq6}
\end{equation}
In single crystals and oriented grains for $H$ applied parallel to
the $c$-axis ($\delta =0$) the magnetization reduces for
sufficiently large anisotropy $\gamma $ to
\begin{equation}
m=-\frac{Qk_{B}TH^{1/2}}{\Phi _{0}^{3/2}}\frac{\gamma
}{z^{1/2}}\frac{dG}{dz},\text{ }z=\frac{H\xi _{ab}^{2}}{\Phi
_{0}},  \label{eq7}
\end{equation}
while in powder samples it reduces to
\begin{equation}
m=-\frac{Qk_{B}\gamma TH^{1/2}}{\Phi _{0}^{3/2}}\left\langle
\frac{\left| \cos \left( \delta \right) \right|
^{3}}{z^{1/2}}\frac{dG}{dz}\right\rangle ,\text{ }z=\frac{H\xi
_{ab}^{2}}{\Phi _{0}}\left| \cos \left( \delta \right) \right| ,
\label{eq8}
\end{equation}
where $\left\langle ...\right\rangle =\int_{0}^{2\pi }...d\delta
$. Thus, when thermal fluctuations dominate and the magnetic field
induced finite size effect sets the limiting length
($L_{H}<L_{I}$), data as shown in Figs.\ref{fig1} and \ref{fig2} \
should collapse on a single curve when plotted as $m/(TH^{1/2})$
\textit{vs.} $t/t_{p}\left( H\right) =\left( T/T_{c}-1\right)
/\left( 1-T_{p}\left( H\right) /T_{c}\right) $ with appropriately
chosen $T_{c}$ and $T_{p}\left( H\right) $. Indeed, according to
Eqs.(\ref{eq1}), (\ref{eq4}), (\ref{eq7}), (\ref{eq8}) and the
critical behavior of the zero field correlation length $\xi
_{ab,c}\left( T\right) =\xi _{ab,c0}\left| t\right| ^{-\nu }$ with
$t=T/T_{c}-1$, the scaling variable $z$ can be expressed as
$z^{-1/2\nu }=a^{1/2\nu }t/t_{p}\left( H\right) $.

\begin{figure}[tbp]
\centering
\includegraphics[totalheight=6cm]{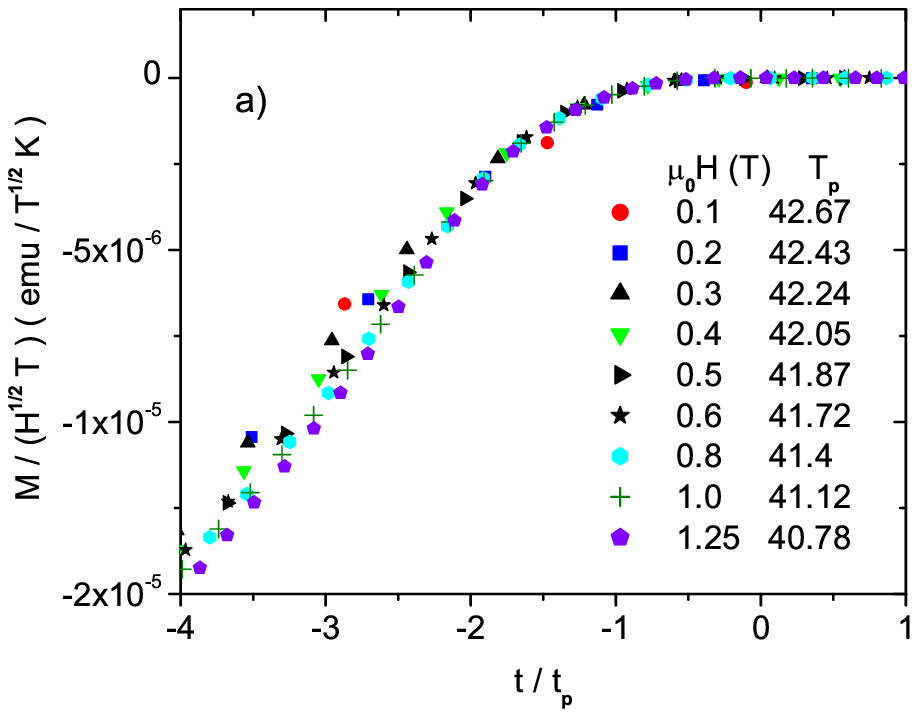}
\includegraphics[totalheight=6cm]{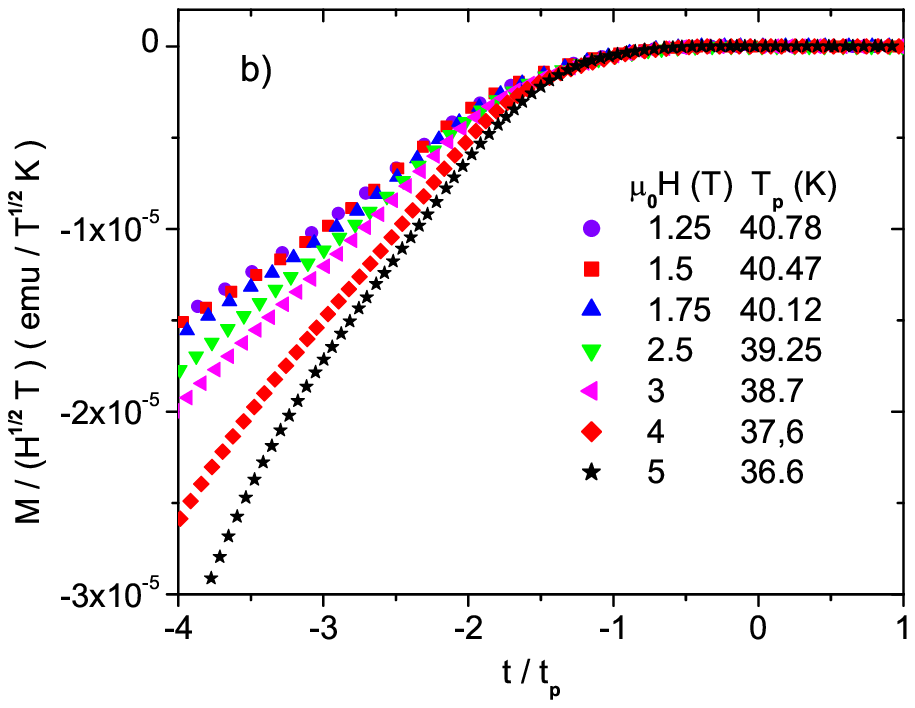}
\caption{$M/(TH^{1/2})$ \emph{vs}. $t/t_{p}\left( H\right) =\left(
T/T_{c}-1\right) /\left( 1-T_{p}\left( H\right) /T_{c}\right) $
for various magnetic fields $H$ with $T_{c}=43$K and the listed
estimates for $T_{p}\left( H\right) $ for the data shown in
Fig.\ref{fig1}. } \label{fig3}
\end{figure}

A glance to Fig.\ref{fig3} shows that for $T_{c}=43$K and the
listed values of $T_{p}\left( H\right) $ the data tend to collapse
around $t/t_{p}\left( H\right) =-1$ on a single curve. However,
with increasing field the range where the data collapse is seen to
shrink. This reflects the fact that the fluctuations of a bulk
superconductor in sufficiently high magnetic fields become
effectively one dimensional, as noted by Lee and Shenoy\cite{lee}.
Here a bulk superconductor behaves like an array of rods parallel
to the magnetic field with diameter $L_{H}$, while the scaling
relation (\ref{eq6}) holds for sufficiently low fields where three
dimensional fluctuations dominate. On the other hand, with
decreasing magnetic field the scaling regime is seen to increase.
Thus, down to $0.1$T the magnetic field appears to set the
limiting length scale so that $L_{I}>L_{H=0.1T}\simeq 814$\AA .

\begin{figure}[tbp]
\centering
\includegraphics[totalheight=6cm]{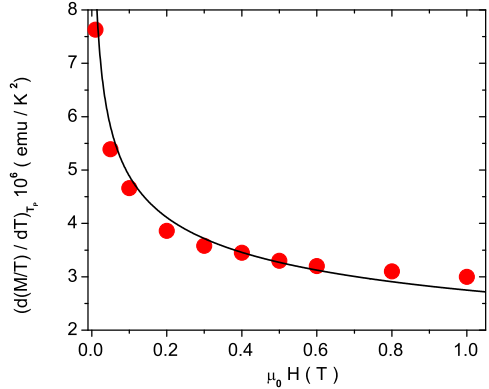}
\caption{$\left. d\left( m/T\right) /dT\right| _{T_{p}}$
\textit{vs}. $H$ derived from the data shown in Fig.\ref{fig1}a.
The solid line is $\left. d\left( m/T\right) /dT\right|
_{T_{p}}=2.75H^{-1/4}$ with $H$ in T, corresponding to
Eq.(\ref{eq9b}) with $\nu =2/3$.} \label{fig4}
\end{figure}

To substantiate the evidence for the dominant role of thermal
fluctuations and to strengthen the discrimination between the
inhomogeneity and magnetic field induced finite size effect
further, we consider the temperature dependence of $d\left(
m/T\right) /dT$. From Eqs.(\ref{eq5}) and (\ref{eq6}) we obtain,

\begin{equation}
\frac{d\left( m/T\right) }{dT}=-\frac{Qk_{B}\varepsilon \gamma
}{\Phi _{0}}\frac{1}{\xi _{ab}^{2}}\frac{d\xi _{ab}}{dT}\left(
-\frac{dG}{dz}+z\frac{d^{2}G}{dz^{2}}\right).   \label{eq9a}
\end{equation}
Due to the magnetic field induced finite size effect $\xi _{ab}$
cannot grow beyond $\xi _{ab}\left( T_{P}\right) =L_{H}$ so that
$d\xi _{ab}/dT=0$ and with that $d\left( m/T\right) /dT=0$ at
$T_{P}$. Indeed, at $T_{P}$ the scaling variable adopts the value
$z_{p}=\epsilon /a$ so that $dG/dz$ and $2zd^{2}G/dz^{2}$ are
temperature and field independent. In powder samples with
sufficiently large anisotropy this occurs at $z_{p}\simeq
\left\langle \epsilon \right\rangle /a=2a/\pi $. Noting that
$\gamma =\xi _{ab0}/\xi _{c0} $ the peak height scales then as
\begin{equation}
\left. \frac{d\left( m/T\right) }{dT}\right| _{T_{p}}\propto
H^{\left( \nu -1\right) /\left( 2\nu \right) }.  \label{eq9b}
\end{equation}
This differs fundamentally from the inhomogeneity induced finite
size effect, where $\left. \left( 1/\xi _{ab}^{2}\right) d\left(
\xi _{ab}\right) /dT\right| _{T_{p}}\propto L_{I}^{1/\nu -1}$ and
$z$ adopts the value $z=\epsilon HL_{I}^{2}/\Phi _{0}$. Thus, the
inhomogeneity induced finite size effect exhibits a weak field
dependence arising from the change of $dG/dz$ and $zdG/dz$. In
Fig.\ref{fig4} we displayed our estimates for $\left. d\left(
m/T\right) /dT\right| _{T_{p}}$ \textit{vs}. $H$. For comparison
we included $\left. d\left( m/T\right) /dT\right| _{T_{p}}\propto
H^{-1/4}$, corresponding to Eq.(\ref{eq9b}) with $\nu =2/3$. The
quantitative agreement for small fields confirms that the limiting
length is set by the magnetic field and not by inhomogeneities in
the grains of our sample. Accordingly, we have shown that in an
applied magnetic field Sr$_{0.9}$La$_{0.1}$CuO$_{2}$ does not
undergo a sharp phase transition from the superconducting to the
normal state up to at least $5$T. Additional stringent features
not accounted for by the mean-field treatment emerge from
Figs.\ref{fig1}b and \ref{fig2}. $dM/dT$ and with that $d\left(
M/T\right) /dT$ are seen to fall below and above $T_{p}$.
Furthermore, this fall is field dependent. According to
Eq.(\ref{eq9a}) the initial fall reflects the temperature
dependence of $\left( 1/\xi _{ab}^{2}\right) d\left( \xi
_{ab}\right) /dT=sgn\left( 1-T/T_{c}\right) \left(
1-T/T_{c}\right) ^{\nu -1}/\left( \nu T_{c}\right) $, while the
field dependence stems from $dG/dz$ $-2zdG/dz$. Thus, the fall of
$dM/dT$ and $d\left( M/T\right) /dT$ below and above $T_{p}$ and
its field dependence confirm the dominant role of 3D-XY
fluctuations, while the rounding of the transition reveals the
magnetic field induced finite size effect. In principle, the
scaling function $G\left( z\right) $ should have a singularity at
some value $z_{m}$ of the scaling variable below $T_{p}\left(
H\right) $ describing the vortex melting transition, but this
singularity is not addressed here.

\begin{figure}[tbp]
\centering
\includegraphics[totalheight=6cm]{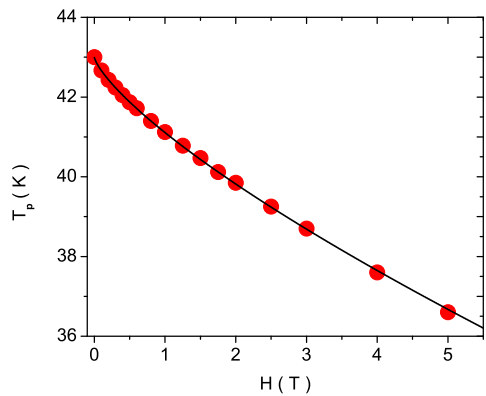}
\caption{$T_{p}$ \textit{vs.} $H$ derived from the scaling plots
shown in Fig.\ref{fig3}. The solid line is Eq.(\ref{eq10}) with
$T_{c}=43$K , $\left( a\xi _{ab0}^{2}/\Phi _{0}\right)
^{3/4}=0.044$, $a=3.12$ and $\nu =2/3$.} \label{fig5}
\end{figure}

In Fig.\ref{fig5} we displayed our estimates for $T_{p}\left(
H\right) $, the temperature where the correlation length equals
the magnetic field induced limiting length scale, as derived from
the scaling plots shown in Fig.\ref{fig3}. According to
Eqs.(\ref{eq1}) and (\ref{eq4}) the leading field dependence is
\begin{equation}
T_{p}\left( H\right) =T_{c}\left( 1-\left( \frac{aH\xi
_{ab0}^{2}}{\Phi _{0}}\right) ^{1/2\nu }\right) .  \label{eq10}
\end{equation}
The solid line in Fig.\ref{fig5} is this relation with $T_{c}=43$K
, $\left( a\xi _{ab0}^{2}/\Phi _{0}\right) ^{3/4}=0.044$, $a=3.12$
and $\nu =2/3$, yielding for the critical amplitude of the
correlation length the estimate $\xi _{ab0}\simeq 32$\AA . With
$\gamma \approx 9$\cite{kim} we obtain for the critical amplitude
of the c-axis correlation length the estimate $\xi _{c0}=\xi
_{ab0}/\gamma \approx 3.6$\AA . The criterion for 3D
superconductivity below $T_{c}$ is $\xi _{c}=\xi _{c0}\left(
1-T/T_{c}\right) ^{-\nu }>c/\sqrt{2}\simeq 2.4$\AA , where
$c\simeq 3.4$\AA\ is the $c$-axis lattice constant. This reveals
that Sr$_{0.9}$La$_{0.1}$CuO$_{2}$ is even at low temperature a 3D
superconductor, as previously suggested \cite{kim}.

\begin{figure}[tbp]
\centering
\includegraphics[totalheight=6cm]{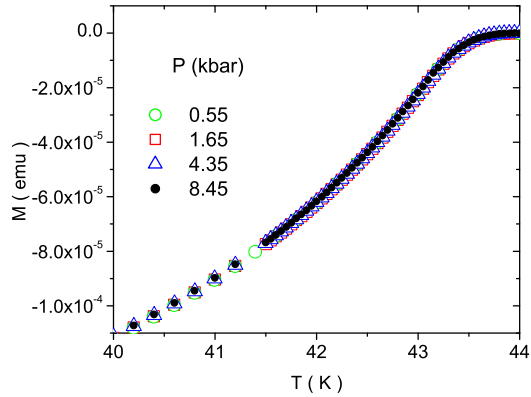}
\caption{ Field cooled ($5$ $10^{-4}$T) magnetization of a
Sr$_{0.9}$La$_{0.1}$CuO$_{2}$ powder sample \textit{vs}. $T$ near
$T_{c}$ for various hydrostatic pressures.} \label{fig6}
\end{figure}

Finally we turn to the pressure effect on the magnetization. Close
to the zero field transition temperature Eq.(\ref{eq8}) reduces
to\cite{jhts,book,parks,bled,tsjh}
\begin{equation}
\frac{M}{VH^{1/2}T_{c}\gamma }=-\frac{QCk_{B}}{\Phi
_{0}^{3/2}},\text{ }C=\left\langle \frac{\left| \cos \left( \delta
\right) \right| ^{3}}{z^{1/2}}\frac{dG}{dz}\right\rangle
_{T\rightarrow T_{c}^{-},H\neq 0}, \label{eq11}
\end{equation}
where $C$ is a universal constant. As the pressure effect on the
magnetization at fixed magnetic field is concerned it implies that
the relative shifts of the magnetization $M$, volume $V$,
anisotropy $\gamma $ and $T_{c}$ are not independent but close to
$T_{c}$ related by
\begin{equation}
\frac{\Delta M}{M}=\frac{\Delta V}{V}+\frac{\Delta \gamma }{\gamma
}+\frac{\Delta T_{c}}{T_{c}}.  \label{eq12}
\end{equation}
On that condition it is impossible to extract these changes from
the temperature dependence of the magnetization. However,
supposing that close to criticality the magnetization data scale
within experimental error as
\begin{equation}
^{i}M\left( T\right) =\text{ }^{j}M\left( aT\right) ,
\label{eq13}
\end{equation}
where $^{i\neq j}M$ denotes the magnetization for different
pressures, the universal relation (\ref{eq12}) reduces to
\begin{equation}
-\frac{\Delta T_{c}}{T_{c}}=\frac{\Delta V}{V}+\frac{\Delta \gamma
}{\gamma }=1-a.  \label{eq14}
\end{equation}
Hence, when Eq.(\ref{eq13}) holds true, the pressure and isotope
effect on $T_{c}$ mirrors that of the anisotropy $\gamma =\xi
_{ab}/\xi _{c}=$ $\lambda _{ab}/\lambda _{c}$ and of the
volume\cite{tsdc}. In Fig.\ref{fig6} we displayed the field cooled
magnetization data for various hydrostatic pressure up to $8.45$
kbar. The hydrostatic pressure was generated in a copper-beryllium
piston cylinder clamp designed for magnetization measurements. The
sample was mixed with Fluorient FC77 (pressure transmitting
medium). The pressure was measured in situ by monitoring the
$T_{c}$ shift of a small piece of In ($T_{c}\left( p=0\right)
\simeq 3.4$K) placed within the cell. Apparently, the data
collapses within experimental error on a single curve\ so that
Eq.(\ref{eq13}) with $a\simeq 1$ applies. On the other hand, given
the bulk modulus $B=1170$ kbar\cite{shaked} there is the volume
change $\Delta V//V\simeq -P/1170$ with $P$ in kbar. It implies
with Eq.(\ref{eq14}) and in the absence of a significant pressure
effect on $T_{c}$ that the volume reduction mirrors essentially
the increase of the anisotropy because
\begin{equation}
\frac{\Delta V}{V}=-\frac{\Delta \gamma }{\gamma }\simeq
-\frac{P\left( \text{kbar}\right) }{1170}.  \label{eq15}
\end{equation}

Together with the experimental fact that in both, electron and
hole doped cuprates, the pressure effect on $T_{c}$ depends
strongly on the dopant concentration and $dT_{c}/dP$ nearly
vanishes close to optimum\cite{murayama}, $\Delta V//V\simeq
-\Delta \gamma /\gamma $ appears to be a generic property of
optimally doped cuprate superconductors. With that it shows that
the anisotropy is essential towards an understanding of
superconductivity in these materials.

We have shown that in Sr$_{0.9}$La$_{0.1}$CuO$_{2}$ thermal
fluctuations do not alter the zero field thermodynamic properties
near $T_{c}$only, but invalidate the assumption of an upper
critical field $H_{c2}$ over a rather extended temperature range.
Indeed, the correlation length $\xi _{ab}$ increases strongly when
$T_{c}$is approached. However, for nonzero magnetic field $H$
there is the limiting length scale $L_{H}=\sqrt{\Phi _{0}/aH}$. It
is comparable to the average distance between vortex lines and
implies that $\xi $ cannot grow beyond $L_{H}$ and with that there
is a magnetic field induced finite size effect. Since $L_{H}$
decreases with increasing field strength it sets the limiting
length scale for sufficiently high fields. In the present case,
both the magnetic field induced shift of the peak location and
peak height in $dM/dT$ clearly revealed that for $\mu _{0}H\gtrsim
0.1$T the magnetic field induced finite size effect allows to
probe the homogeneous parts of the grains with dimension $814$\AA
. As a result, whenever the thermal fluctuation dominated regime
is accessible, homogeneous type II superconductors do not undergo
in nonzero magnetic field a continuous phase transition,
\textit{e.g.} to a state with zero resistance. Furthermore, we
have shown that the pressure effect on the magnetization does not
provide estimates for the shift of the transition temperature
only, but involves the change of the volume and anisotropy as
well. Consistent with the generic behavior of optimally doped
cuprates we found that in Sr$_{0.9}$La$_{0.1}$CuO$_{2}$ the
pressure effect on $T_{c}$ is negligibly small, while the negative
value of the relative volume change mirrors that of the
anisotropy.

\acknowledgments This work was partially supported by the Swiss
National Science Foundation, the NCCR program {\it Materials with
Novel Electronic Properties} (MaNEP) sponsored by the Swiss
National Science Foundation and the Creative Research Initiatives
of the Korean Ministry of Science and Technology.

\end{document}